# Adiabatic quantum-flux-parametron with delay-line clocking: logic gate demonstration and phase skipping operation

Taiki Yamae,[1,2] Naoki Takeuchi,[3,4,*] and Nobuyuki Yoshikawa[1,4]

[1] Department of Electrical and Computer Engineering, Yokohama National University, 79-5 Tokiwadai, Hodogaya, Yokohama 240-8501, Japan

[2] Research Fellow of Japan Society for the Promotion of Science, 5-3-1 Kojimachi, Chiyoda, Tokyo 102-0083, Japan

[3] Research Center for Emerging Computing Technologies, National Institute of Advanced Industrial Science and Technology (AIST), 1-1-1 Umezono, Tsukuba 305-8568, Japan

[4] Institute of Advanced Sciences, Yokohama National University, 79-5 Tokiwadai, Hodogaya, Yokohama 240-8501, Japan

[*] n-takeuchi@aist.go.jp

**Abstract.** Adiabatic quantum-flux-parametron (AQFP) logic is an energy-efficient superconductor logic family. The latency of AQFP circuits is relatively long compared to that of other superconductor logic families and thus such circuits require low-latency clocking schemes. In a previous study, we proposed a low-latency clocking scheme called delay-line clocking, in which the latency for each logic operation is determined by the propagation delay of the excitation current, and demonstrated a simple AQFP buffer chain that adopts delay-line clocking. However, it is unclear whether more complex AQFP circuits can adopt delay-line clocking. In the present study, we demonstrate AQFP logic gates (AND and XOR gates) that use delay-line clocking as a step towards implementing large-scale AQFP circuits with delay-line clocking.

AND and XOR gates with a latency of approximately 20 ps per gate are shown to operate at up to 5 and 4 GHz, respectively, in experiments. We also demonstrate that delay-line clocking enables phase skipping operation, in which some of the AQFP buffers for phase synchronization are removed to reduce the junction count and energy dissipation. The results indicate that delay-line clocking can enable low-latency, low-energy, large-scale AQFP circuits.

## 1. Introduction

Josephson junctions can be used as superconducting switching devices with a few-picosecond characteristic time and a sub-aJ energy barrier height. Superconducting digital circuits based on Josephson junctions [1–4] can thus achieve high operating speed and low switching energy, making them a key technology for developing superconducting information systems, such as microprocessors [5–8], neuromorphic computing systems [9, 10], and reversible computers [11, 12]. To realize large-scale, high-performance superconducting systems, fabrication processes have been continuously improved [13–15] and many types of circuit design tool have been developed [16–18].

Adiabatic quantum-flux-parametron (AQFP) logic [19] is an energy-efficient superconductor logic family based on the quantum flux parametron [20, 21]. Because adiabatic switching [22, 23] enables a logic gate to switch without rapid state transition, the switching energy of an AQFP gate can be reduced to less than the Josephson energy [24]. Recently, we designed and demonstrated the 4-bit Monolithic Adiabatic iNtegration Architecture (MANA) microprocessor that uses AQFP logic [8], thereby proving the high robustness and high energy efficiency of AQFP circuits. We found that the latency of this microprocessor is relatively long (for instance, 5400 ps for a 5-GHz clock frequency). The reason for this long latency is that the microprocessor is driven by four-phase clocking [25, 26], in which AQFP circuits are driven by a pair of excitation currents with a phase separation of 90° and a latency of a quarter clock cycle (50 ps at 5 GHz) per gate is required. To reduce the latency of AQFP microprocessors, it is necessary to adopt low-latency clocking schemes such as delay-line clocking [27], in which the latency of each logic operation is determined by the propagation delay of the excitation current. In a previous study, we demonstrated an AQFP buffer chain with a latency of approximately 10 ps per gate achieved using delay-line clocking [27]. This suggests that the latency of large-scale AQFP circuits, such as microprocessors, can be significantly reduced by adopting delay-line clocking.

In the present study, we conduct a simulation and experiments as a step toward implementing large-scale AQFP circuits with delay-line clocking. Specifically, we show that (i) delay-line clocking can be used to operate AQFP logic gates with low latency and that (ii) delay-line clocking can reduce the number of buffers required for phase synchronization. Because delay-line clocking adoption has been limited to buffers [27], we first demonstrate AQFP logic gates (AND and XOR gates) that use delay-line clocking to show that delay-line clocking is applicable to complex AQFP circuits. Then, we show that delay-line clocking can remove some buffers required for phase synchronization, i.e., it enables the skipping of some excitation phases. This is an important advantage of delay-line clocking because large-scale AQFP circuits typically require many additional buffers for phase synchronization [28]. Finally, we fabricate and operate a test circuit to demonstrate these properties at high clock frequencies.

## 2. Logic gate demonstration

Figure 1(a) shows a schematic diagram of an AQFP circuit that adopts delay-line clocking for numerical simulation, where the circuit under test (CUT) is surrounded by peripheral AQFP buffers that prevent unwanted interaction with the input ports and output termination. Here, the CUT is an XOR gate. The entire circuit is powered and clocked by the excitation current $I_x$, which applies an ac magnetic flux with an amplitude of $0.5\Phi_0$ to each gate, where $\Phi_0$ is the flux quantum. The dc offset current $I_d$ applies a dc magnetic flux of $0.5\Phi_0$ to each gate. $I_x$ and $I_d$ flow through the AQFP gates and delay lines, which are 50-$\Omega$ transmission lines, and are terminated by 50-$\Omega$ resistors. In this way, logic operations are performed at the rising edge of $I_x$ with a latency of $T$ per gate, where $T$ is the propagation delay of each delay line. In the numerical simulation, the device parameters are based on the AIST 10 kA cm$^{-2}$ Nb high-speed standard process (HSTP) [26]. The numerical simulation is conducted using the Josephson circuit simulator JSIM [29]. Figure 1(b) shows transient analysis results for an XOR gate at 5 GHz for $T = 10$ ps, where $I_{sta}$, $I_{stb}$, and $I_{stq}$ are the signal currents that represent the logic states of buffers A, B, and Q in figure 1(a), respectively. Note that the XOR gate is composed of splitters, AND gates, and an OR gate, and thus requires three excitation phases. Hence, seven excitation phases are included between gates A (B) and Q, so that the latency between $I_{sta}$ ($I_{stb}$) and $I_{stq}$ is 70 ps. This figure shows that the XOR gate operates correctly with a latency of 10 ps per gate.

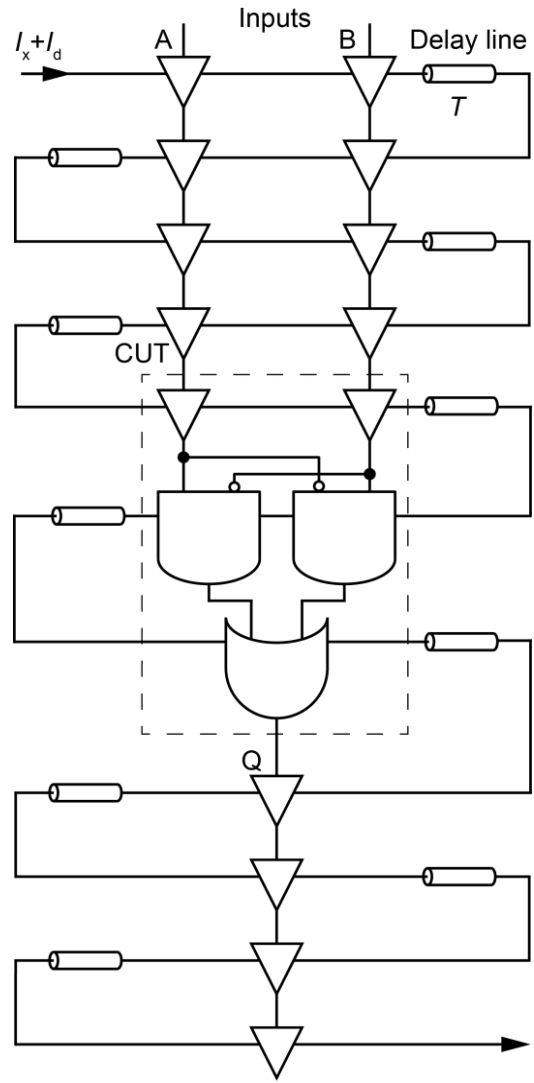

(a)

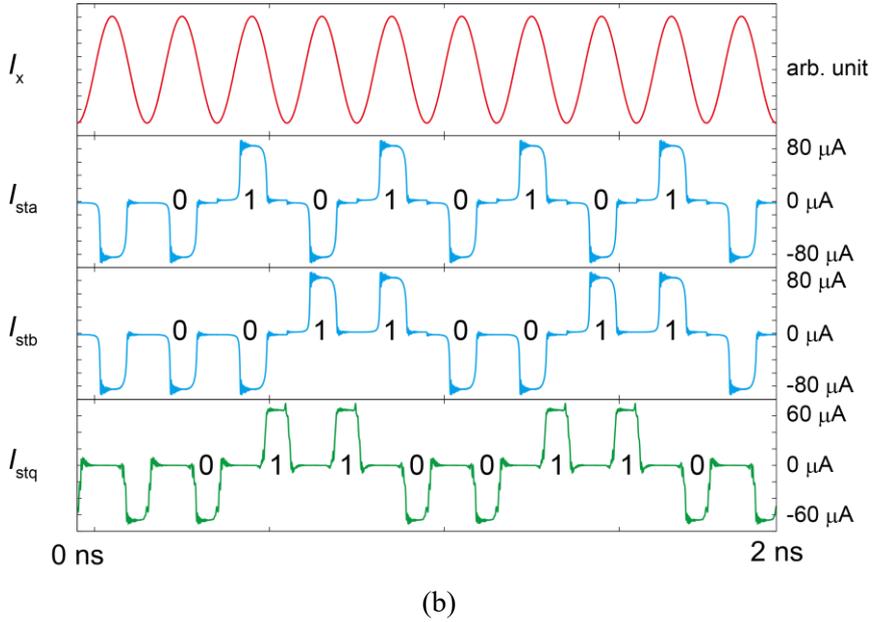

**Figure 1.** (a) Schematic diagram of an AQFP circuit that uses delay-line clocking for numerical simulation. (b) Transient analysis results for an XOR gate operating at 5 GHz for $T$ = 10 ps.

We investigated the operating margins for an AND gate and an XOR gate that use delay-line clocking to show that AQFP logic gates can operate with wide operating margins for small $T$. Figure 2 shows numerical simulation results of the operating margins in terms of excitation power $P_x$ (i.e., the power of $I_x$) applied to AND and XOR gates as a function of $T$ at 5 GHz. The results for a buffer are also shown for comparison. The enclosed regions represent the operation regions of the logic gates. Each logic gate was simulated using the schematic shown in figure 1(a). Figure 2 shows that both AND and XOR gates operate with wide operating margins for small $T$, as is the case for a buffer. Of note, the XOR gate achieves wide operating margins even though it is composed of multiple logic gates (i.e., splitters, AND gates, and an OR gate) that comprise different building blocks [30]. This suggests that delay-line clocking is applicable to complex AQFP circuits.

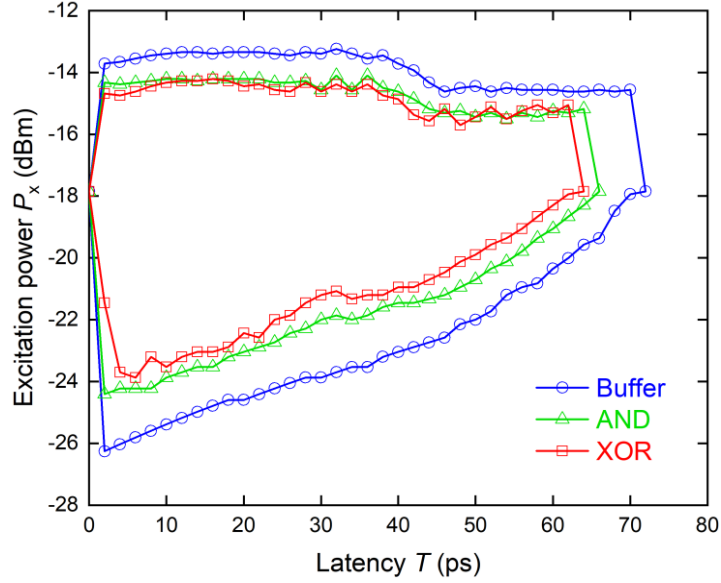

**Figure 2.** Simulation results of operating margins for a buffer, an AND gate, and an XOR gate that use delay-line clocking as a function of $T$ at 5 GHz.

We also calculated the energy dissipation of AND and XOR gates to show that AQFP logic gates that adopt delay-line clocking can achieve both low energy dissipation and low latency. Figure 3 shows the numerical simulation results of the energy dissipation per operation $E$ for AND and XOR gates with delay-line clocking as a function of $T$ for an operating frequency of 5 GHz. The results for a buffer are also shown. The energy dissipation was calculated by integrating the product of the excitation current and voltage across the excitation inductor in each gate over time; more details can be found in the literature [31]. Because the energy dissipation of AQFP circuits depends on input data combinations [32], the energy dissipation for AND and XOR gates shown in figure 3 is the average among all input data combinations. Figure 3 shows that $E$ does not change significantly for a $T$ value of between 10 and 50 ps for both AND and XOR gates, indicating that AQFP logic gates that adopt delay-line clocking can operate with small energy dissipation and a latency that is much less than that for conventional four-phase clocking (50 ps at 5 GHz). For a $T$ value of less than 10 ps, $E$ increases significantly as $T$ decreases for all logic gates. This is because for very small $T$, each AQFP gate switches before the previous gate is fully excited and the amplitude of the input current becomes sufficiently large [27].

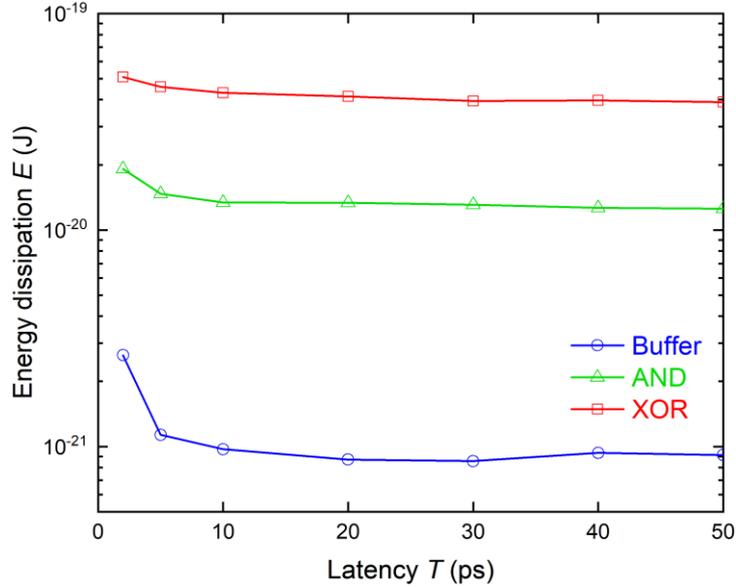

**Figure 3.** Simulation results of energy dissipation for a buffer, an AND gate, and an XOR gate that use delay-line clocking as a function of $T$ at 5 GHz.

## 3. Phase skipping operation

In general, most logic gates in a large-scale AQFP circuit are buffers [28] because a buffer must be inserted to transmit data from an excitation phase to the next excitation phase for four-phase clocking [26]; if this were not done, a gate that transmits data would reset before the next gate is excited, thereby losing the data. For instance, in our previous 4-bit microprocessor [8], approximately 70% of the Josephson junctions are used for buffers. Thus, to further reduce energy dissipation, it is necessary to reduce the number of buffers in the circuit design. An advantage of delay-line clocking is that it is not always necessary to insert a buffer for data transmission because the latency, or clock skew, between adjacent excitation phases (i.e., $T$) can be much shorter than that for four-phase clocking (i.e., a quarter clock cycle), so that in some cases data can be transmitted across excitation phases without buffers [33]. This type of data transmission is hereafter referred to as phase skipping operation. Figure 4(a) shows a schematic diagram of AQFP buffer chains for demonstrating phase skipping operation. Three types of buffer chain are shown, namely a conventional buffer chain without phase skipping operation, a 1-phase skipping buffer chain, and a 2-phase skipping buffer chain. Here, $\phi_1$ through $\phi_5$ denote excitation phases. In the 1-phase skipping buffer chain, a buffer is not placed in the third excitation phase $\phi_3$; thus, data are transmitted directly from the buffer in the second phase $\phi_2$ to that in the fourth phase $\phi_4$, across $\phi_3$. Note that the latency between the buffers in $\phi_2$ and $\phi_4$ is $2T$, so that 1-phase skipping must be conducted with $2T$ that is smaller than the maximum allowable latency $T_{max}$. Similarly, 2-phase skipping must be conducted with $3T$ that is smaller than $T_{max}$. For instance, $T_{max}$ is approximately 70 ps at 5 GHz for a buffer, as shown in figure 2. In this case, 1-phase skipping can be conducted when $2T$ is less than approximately 70 ps, i.e., $T \lesssim 35$ ps. Figure 4(b) shows transient analysis results for the 1-phase skipping buffer chain at 5 GHz for $T$ = 10 ps, where $I_{st1}$ through $I_{st5}$ are the signal currents of the buffers in $\phi_1$ through $\phi_5$, respectively. This figure shows that although a buffer is not placed in $\phi_3$, the 1-phase skipping buffer chain operates correctly because the latency between $\phi_2$ and $\phi_4$ ($2T$ = 20 ps) is much smaller than $T_{max}$.

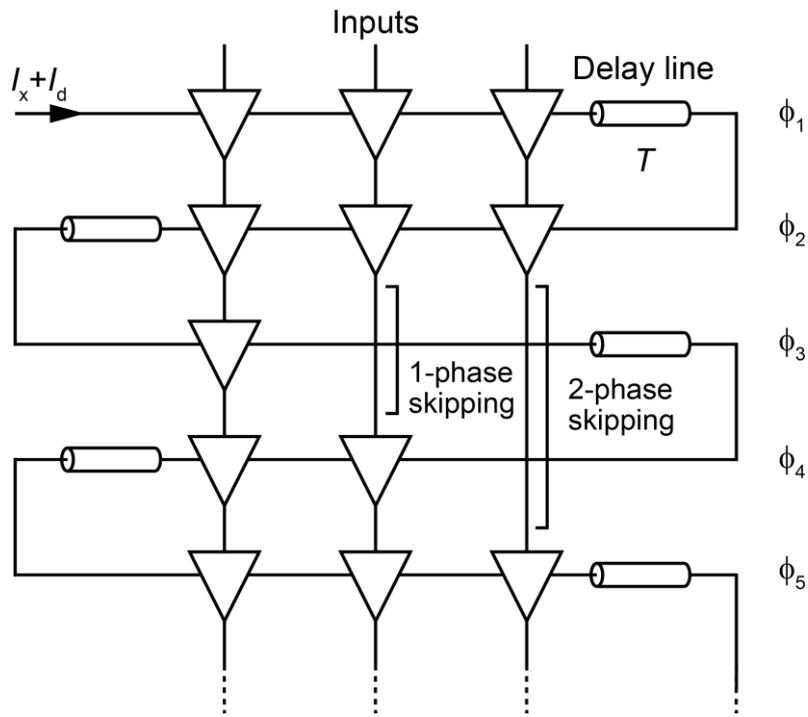

(a)

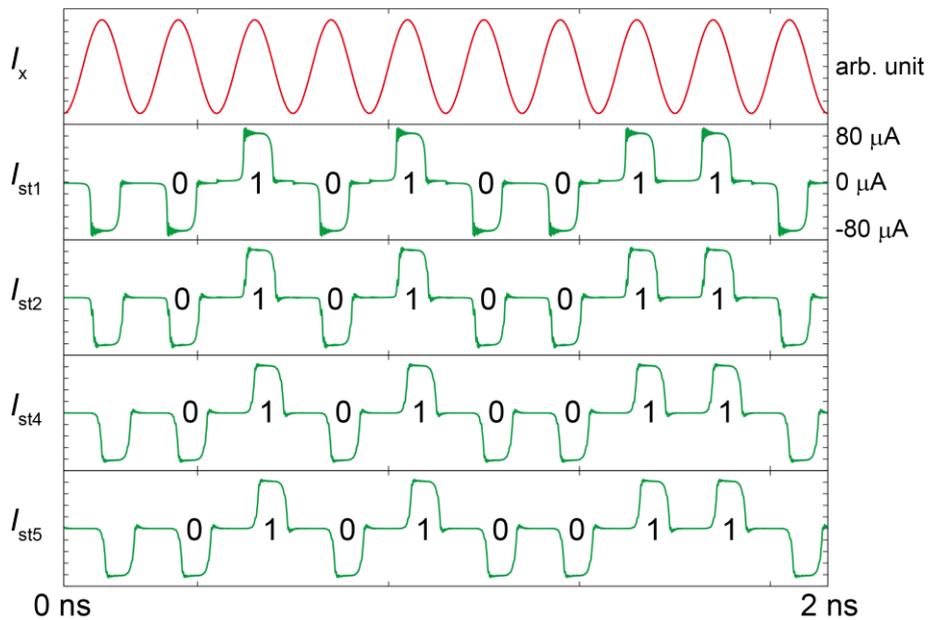

(b)

**Figure 4.** (a) Schematic diagram of AQFP buffer chains for demonstrating phase skipping operation with delay-line clocking. (b) Transient analysis results for the 1-phase skipping buffer chain at 5 GHz for $T = 10$ ps.

We investigated the operating margins for phase skipping buffer chains to determine the relation between latency and the number of excitation phases that a buffer can skip. Figure 5 shows the numerical simulation results of the operating margins in terms of the excitation power $P_x$ applied to phase skipping buffer chains (0-, 1-, 2-, 3-, and 4-phase skipping buffer chains) as a function of $T$ for an operating frequency of 5 GHz. The enclosed regions represent the operation regions of the phase skipping buffer chains. These simulation results indicate that the number of excitation phases that a buffer can skip increases as $T$ decreases. For example, 0-, 1-, and 2-phase skipping operations are possible for $T = 20$ ps, and 0-, 1-, 2-, 3-, and 4-phase skipping operations are possible for $T = 10$ ps.

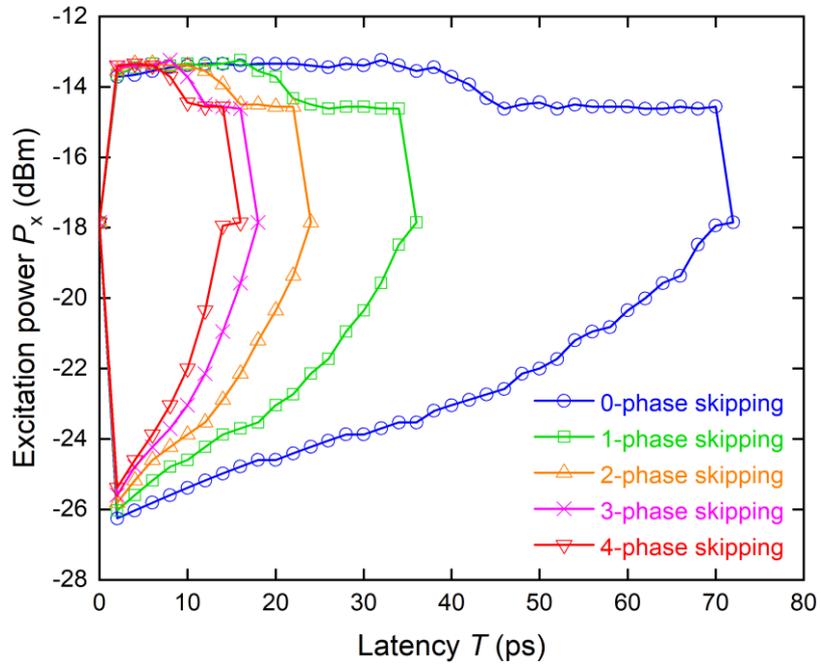

**Figure 5.** Simulation results of operating margins for 0-, 1-, 2-, 3-, and 4-phase skipping buffer chains that use delay-line clocking as a function of $T$ at 5 GHz.

## 4. Measurement results

Figure 6(a) shows a micrograph of AQFP circuits for demonstrating logic gates and phase skipping operation with delay-line clocking. The circuits include an AND gate, an XOR gate, and phase skipping buffer chains (0-, 1-, 2-, 3-, and 4-phase skipping buffer chains). Figure 6(b) shows a close-up view of the phase skipping buffer chains. The circuits were designed and fabricated using the HSTP. The characteristic impedance of the excitation lines in the AQFP gates was designed to be 50 Ω using InductEx [34], which is a 3D simulator for extracting circuit parameters for superconducting integrated circuits, to allow the microwave excitation current to propagate without reflections [35]. These circuits were powered and clocked by a sinusoidal excitation current $I_x$ with a dc offset current $I_d$. $I_x$ and $I_d$ flow through 50-Ω delay-lines and are terminated by off-chip 50-Ω terminators [24]. In the present study, $T$ was designed to be approximately 20 ps for safety. $I_x$ was supplied by a signal generator (Anritsu, MG3710A) and the input currents were supplied by a pattern generator (Agilent, N4906B). The output voltages of the circuits were amplified by voltage drivers [36] that use stacked dc superconducting quantum interference devices. During the experiments, the circuits were immersed in liquid He at 4.2 K.

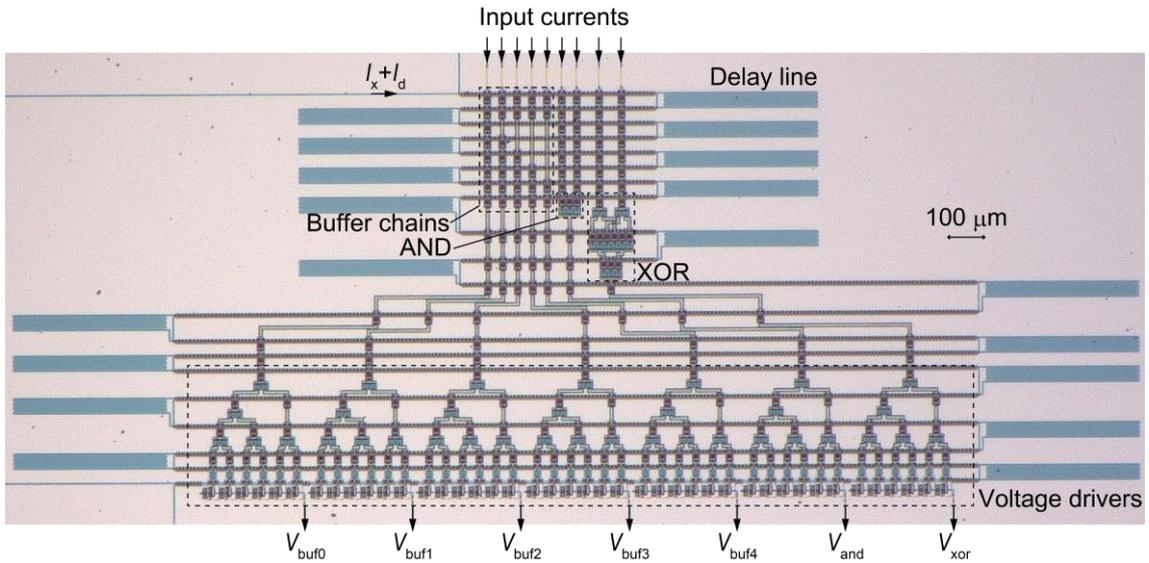

(a)

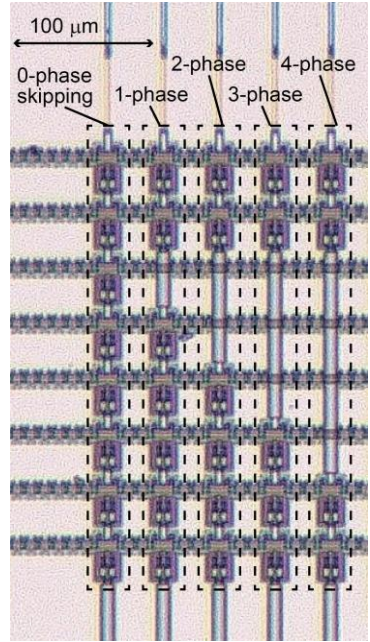

(b)

**Figure 6.** Micrograph of AQFP circuits. (a) AQFP circuits that include phase skipping buffer chains, an AND gate, and an XOR gate. (b) Close-up view of the phase skipping buffer chains.

Figure 7 shows the measurement waveforms of the XOR gate at 3 GHz, where $I_{inb}$ is the pseudorandom binary input current applied to input B. $V_{xor,0}$ ($V_{xor,1}$) is the output voltage when input A is fixed to 0 (1) by a dc signal current. The output should be equal to B for A = 0 and equal to ¬B for A = 1, which was confirmed by an error detector (Agilent, N4906B). Note that this cannot be confirmed by the waveforms shown in figure 7 because the latency between $I_{inb}$ and $V_{xor,0}$ or $V_{xor,1}$ is large (~27 ns) due to the propagation delay in the dipping probe and cables. $V_{xor,1}$ is the inversion of $V_{xor,0}$, which shows that the XOR gate operates correctly. The measured operating margin, in which bit error rates are less than $10^{-5}$, for $P_x$ at 3 GHz was 5.9 dB (−21.9 to −16.0 dBm). For the AND gate at 4 GHz, we found an operating margin for $P_x$ of 4.4 dB (−20.1 to −15.7 dBm). Figure 8 shows the operating margins in terms of the excitation power $P_x$ applied to the AND and XOR gates as a function of operating frequency $f$. The results for a buffer chain are also shown for comparison. We found that the AND and XOR gates can operate up to 5 and 4 GHz, respectively. These results indicate that AQFP logic gates with delay-line clocking can operate at frequencies in the gigahertz range. Unfortunately, the AND and XOR gates could not operate at 6 GHz, at which the output waveforms were unstable and the bit error rates were high. A possible reason for this is that the bandwidth of the measurement setup (e.g., dipping probe) caused unwanted reflections in the excitation current.

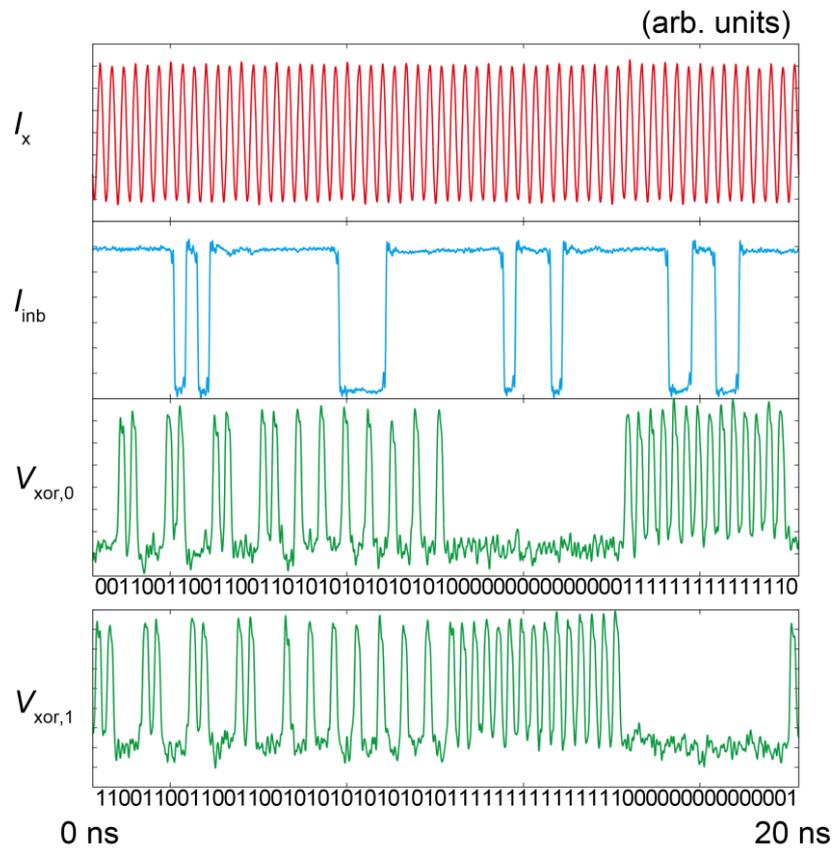

**Figure 7.** Measurement waveforms obtained at 3 GHz for the XOR gate.

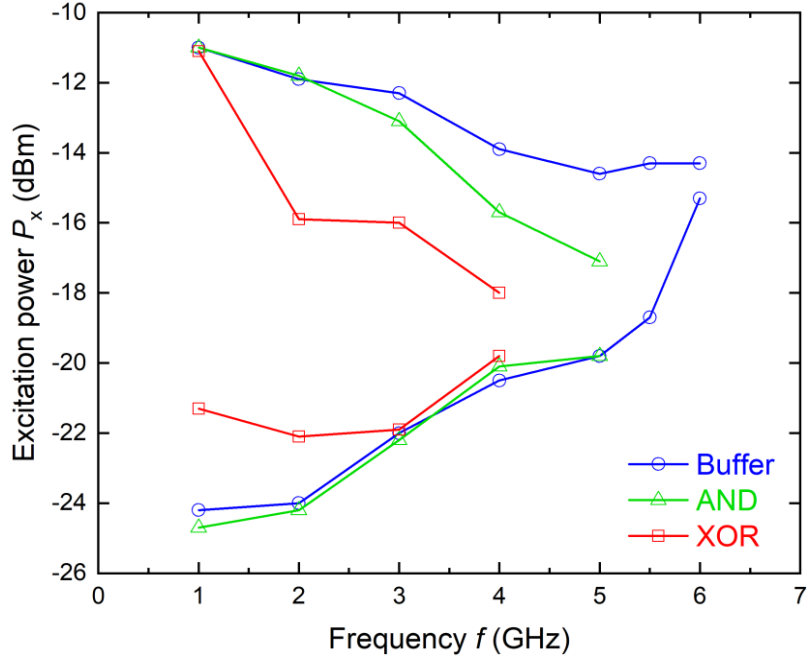

**Figure 8.** Measurement results of operating margins for a buffer, an AND gate, and an XOR gate as a function of operating frequency.

Figure 9 shows the waveforms of the 0-phase skipping buffer chain operating at 5.5 GHz, where $I_{in0}$ is the pseudorandom binary input current. As with the case of the XOR gate, the relation between the input ($I_{in0}$) and output ($V_{buf0}$) was analyzed by an error detector, whereas this relation cannot be confirmed by the waveforms shown in figure 9 because of the latency between $I_{in0}$ and $V_{buf0}$. The measured operating margin for $P_x$ was 4.4 dB (−18.7 to −14.3 dBm). Figure 10 shows the operating margins in terms of the excitation power $P_x$ applied to the phase skipping buffer chains as a function of $f$. We tested each buffer chain up to 6 GHz. We found that the 0-phase skipping buffer chain can operate at up to 6 GHz, the 1- and 2-phase skipping buffer chains can operate at up to 5.5 GHz, and the 3- and 4-phase skipping buffer chains can operate at up to 4.5 and 3.5 GHz, respectively. These latter values agree with the maximum operating frequencies estimated in the numerical simulation (approximately 4.4 and 3.5 GHz for the 3- and 4-phase skipping buffer chains, respectively). The maximum operating frequency decreases with increasing number of skipped phases because the clock skew between adjacent buffers increases with the number of skipped phases and thus exceeds $T_{max}$ at some point.

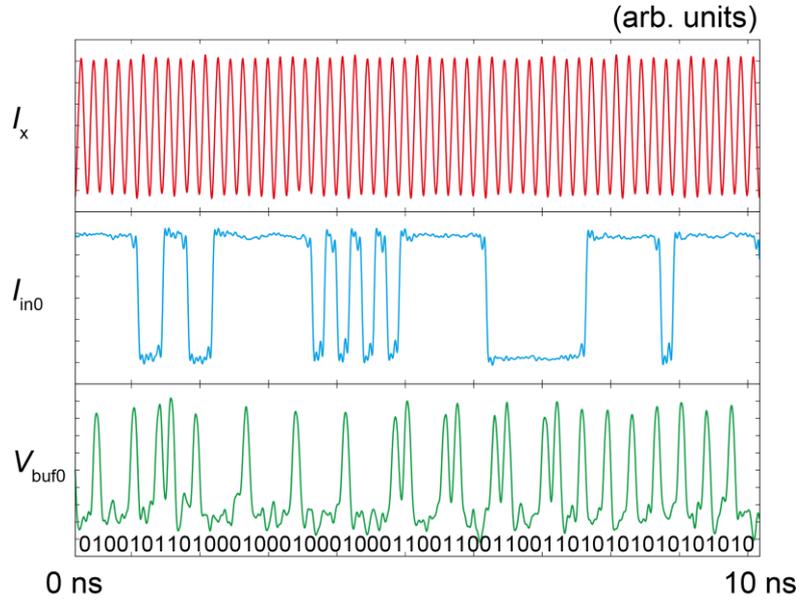

**Figure 9.** Measurement waveforms obtained at 5.5 GHz for a buffer chain.

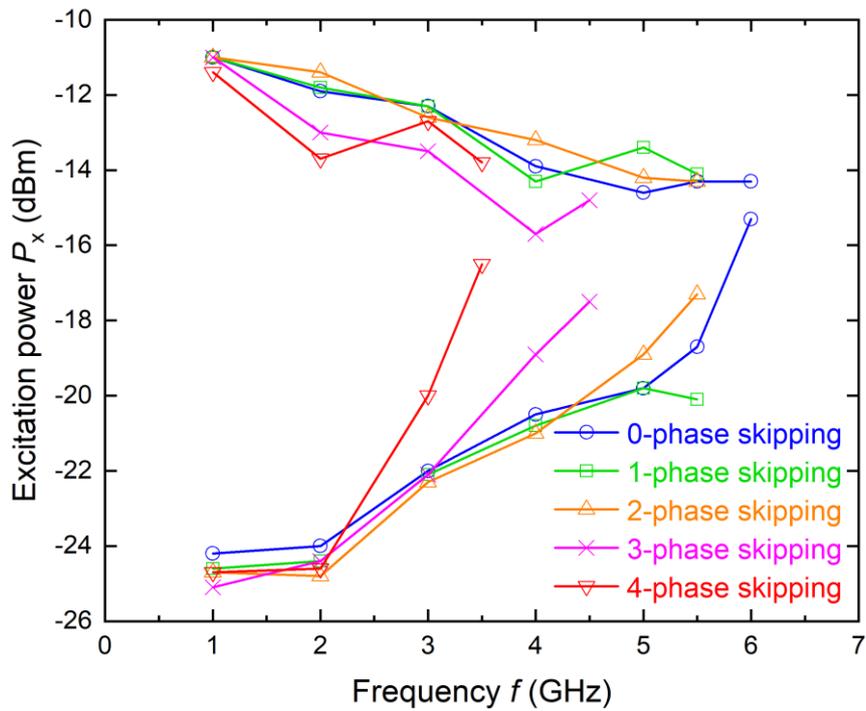

**Figure 10.** Measurement results of operating margins for phase skipping buffer chains as a function of operating frequency.

## 5. Conclusion

We demonstrated AND and XOR gates that use delay-line clocking at clock frequencies in the gigahertz range to show that delay-line clocking is applicable to complex AQFP circuits. We also demonstrated phase skipping operation to show that some of the buffers for phase synchronization can be removed when using delay-line clocking, which results in a reduction in the junction count and energy dissipation. Our simulation and measurement results indicate that delay-line clocking can reduce both the latency and energy dissipation of large-scale AQFP circuits. To clarify the advantage of delay-line clocking, here we compare AQFP logic using delay-line clocking with that using four-phase clocking and another superconductor logic, rapid single-flux-quantum (RSFQ) logic [1]. Specifically, we compare the energy-delay product (EDP), i.e., the product of energy dissipation per clock cycle and gate delay, of each logic. The EDP of an AQFP gate using delay-line clocking is 2.8 zJ × 10 ps = $2.8 \times 10^{-32}$ J·s, and that using four-phase clocking is 2.8 zJ × 50 ps = $1.4 \times 10^{-31}$ J·s, where the energy dissipation is based on our previous study [24]. The EDP of an RSFQ gate is 17 aJ × 4 ps = $6.8 \times 10^{-29}$ J·s, which was estimated from the literature [37] for an operating frequency of 50 GHz. The above comparison indicates the extremely high energy efficiency of AQFP logic using delay-line clocking.

## Acknowledgements

The present study was supported by JSPS KAKENHI (Grants No. 18H01493, No. 18H05245, No. 19H05614, and No. 20J20495). The devices were fabricated in the clean room for analog-digital superconductivity (CRAVITY) of the National Institute of Advanced Industrial Science and Technology (AIST). We would like to thank C. J. Fourie for providing a 3D inductance extractor, InductEx, and H. Suzuki for supporting measurements.

# References

[1] Likharev K K and Semenov V K 1991 RSFQ logic/memory family: a new Josephson-junction technology for sub-terahertz-clock-frequency digital systems *IEEE Trans. Appl. Supercond.* **1** 3–28

[2] Mukhanov O A 2011 Energy-Efficient Single Flux Quantum Technology *IEEE Trans. Appl. Supercond.* **21** 760–9

[3] Herr Q P, Herr A Y, Oberg O T and Ioannidis A G 2011 Ultra-low-power superconductor logic *J. Appl. Phys.* **109** 103903

[4] Tanaka M, Ito M, Kitayama A, Kouketsu T and Fujimaki A 2012 18-GHz, 4.0-aJ/bit Operation of Ultra-Low-Energy Rapid Single-Flux-Quantum Shift Registers *Jpn. J. Appl. Phys.* **51** 053102

[5] Kashima R, Nagaoka I, Tanaka M, Yamashita T and Fujimaki A 2021 64-GHz Datapath Demonstration for Bit-Parallel SFQ Microprocessors Based on a Gate-Level-Pipeline Structure *IEEE Trans. Appl. Supercond.* **31** 1301006

[6] Herr A Y, Herr Q P, Oberg O T, Naaman O, Przybysz J X, Borodulin P and Shauck S B 2013 An 8-bit carry look-ahead adder with 150 ps latency and sub-microwatt power dissipation at 10 GHz *J. Appl. Phys.* **113** 033911

[7] Kirichenko A F, Vernik I V, Kamkar M Y, Walter J, Miller M, Albu L R and Mukhanov O A 2019 ERSFQ 8-Bit Parallel Arithmetic Logic Unit *IEEE Trans. Appl. Supercond.* **29** 1302407

[8] Ayala C L, Tanaka T, Saito R, Nozoe M, Takeuchi N and Yoshikawa N 2021 MANA: A Monolithic Adiabatic iNtegration Architecture Microprocessor Using 1.4-zJ/op Unshunted Superconductor Josephson Junction Devices *IEEE J. Solid-State Circuits* **56** 1152–65

[9] Bozbey A, Karamuftuoglu M A, Razmkhah S and Ozbayoglu M 2018 Single Flux Quantum Based Ultrahigh Speed Spiking Neuromorphic Processor Architecture *arXiv*:1812.10354

[10] Schneider M L, Donnelly C A, Haygood I W, Wynn A, Russek S E, Castellanos-Beltran M A, Dresselhaus P D, Hopkins P F, Pufall M R and Rippard W H 2020 Synaptic weighting in single flux quantum neuromorphic computing *Sci. Rep.* **10** 934

[11] Frank M P, Lewis R M, Missert N A, Wolak M A and Henry M D 2019 Asynchronous Ballistic Reversible Fluxon Logic *IEEE Trans. Appl. Supercond.* **29** 1302007

[12] Osborn K D and Wustmann W 2021 Reversible Fluxon Logic With Optimized CNOT Gate Components *IEEE Trans. Appl. Supercond.* **31** 1300213

[13] Tolpygo S K, Bolkhovsky V, Rastogi R, Zarr S, Day A L, Golden E, Weir T J, Wynn A and Johnson L M 2019 Advanced Fabrication Processes for Superconductor Electronics: Current Status and New Developments *IEEE Trans. Appl. Supercond.* **29** 1102513

[14] Ying L, Zhang X, Niu M, Ren J, Peng W, Maezawa M and Wang Z 2021 Development of Multi-Layer Fabrication Process for SFQ Large Scale Integrated Digital Circuits *IEEE Trans. Appl. Supercond.* **31** 1301504

[15] Hidaka M and Nagasawa S 2021 Fabrication Process for Superconducting Digital Circuits *IEICE Trans. Electron.* **E104-C** 405–10

[16] Fourie C J *et al* 2019 ColdFlux Superconducting EDA and TCAD Tools Project: Overview and Progress *IEEE Trans. Appl. Supercond.* **29** 1300407

[17] Razmkhah S and Febvre P 2020 JOINUS: A User-Friendly Open-Source Software to Simulate Digital Superconductor Circuits *IEEE Trans. Appl. Supercond.* **30** 1300807

[18] Inamdar A *et al* 2021 Development of Superconductor Advanced Integrated Circuit Design Flow Using Synopsys Tools *IEEE Trans. Appl. Supercond.* **31** 1301907

[19] Takeuchi N, Ozawa D, Yamanashi Y and Yoshikawa N 2013 An adiabatic quantum flux parametron as an ultra-low-power logic device *Supercond. Sci. Technol.* **26** 035010

[20] Loe K and Goto E 1985 Analysis of flux input and output Josephson pair device *IEEE Trans. Magn.* **21** 884–7

[21] Hosoya M, Hioe W, Casas J, Kamikawai R, Harada Y, Wada Y, Nakane H, Suda R and Goto E 1991 Quantum flux parametron: a single quantum flux device for Josephson supercomputer *IEEE Trans. Appl. Supercond.* **1** 77–89

[22] Likharev K K 1977 Dynamics of some single flux quantum devices: I. Parametric quantron *IEEE Trans. Magn.* **13** 242–4

[23] Koller J G and Athas W C 1992 Adiabatic switching, low energy computing, and the physics of storing and erasing information *Workshop Physics Computing* (IEEE) pp 267–70

[24] Takeuchi N, Yamae T, Ayala C L, Suzuki H and Yoshikawa N 2019 An adiabatic superconductor 8-bit adder with $24k_BT$ energy dissipation per junction *Appl. Phys. Lett.* **114** 042602

[25] Hioe W, Hosoya M, Kominami S, Yamada H, Mita R and Takagi K 1995 Design and operation of a Quantum Flux Parametron bit-slice ALU *IEEE Trans. Appl. Supercond.* **5** 2992–5

[26] Takeuchi N, Nagasawa S, China F, Ando T, Hidaka M, Yamanashi Y and Yoshikawa N 2017 Adiabatic quantum-flux-parametron cell library designed using a 10 kA cm$^{-2}$ niobium fabrication process *Supercond. Sci. Technol.* **30** 035002

[27] Takeuchi N, Nozoe M, He Y and Yoshikawa N 2019 Low-latency adiabatic superconductor logic using delay-line clocking *Appl. Phys. Lett.* **115** 072601

[28] Chen O, Cai R, Wang Y, Ke F, Yamae T, Saito R, Takeuchi N and Yoshikawa N 2019 Adiabatic Quantum-Flux-Parametron: Towards Building Extremely Energy-Efficient Circuits and Systems *Sci. Rep.* **9** 10514

[29] Fang E S and Van Duzer T 1989 A Josephson integrated circuit simulator (JSIM) for superconductive electronics application Extended Abstracts of Int. Superconductivity Electronics Conf. (ISEC) (Berkeley, CA: UC Berkeley) pp 407–10

[30] Takeuchi N, Yamanashi Y and Yoshikawa N 2015 Adiabatic quantum-flux-parametron cell library adopting minimalist design *J. Appl. Phys.* **117** 173912

[31] Yamae T, Takeuchi N and Yoshikawa N 2019 Systematic method to evaluate energy dissipation in adiabatic quantum-flux-parametron logic *J. Appl. Phys.* **126** 173903

[32] Takeuchi N, Yamanashi Y and Yoshikawa N 2017 Reversibility and energy dissipation in adiabatic superconductor logic *Sci. Rep.* **7** 75

[33] Saito R, Ayala C L and Yoshikawa N 2021 Buffer Reduction via N-Phase Clocking in Adiabatic Quantum-Flux-Parametron Benchmark Circuits *IEEE Trans. Appl. Supercond.* **31** 1302808

[34] Fourie C J 2015 Full-Gate Verification of Superconducting Integrated Circuit Layouts With InductEx *IEEE Trans. Appl. Supercond.* **25** 1300209

[35] Takeuchi N, Suzuki H, Fourie C J and Yoshikawa N 2021 Impedance Design of Excitation Lines in Adiabatic Quantum-Flux-Parametron Logic Using InductEx *IEEE Trans. Appl. Supercond.* **31** 1300605

[36] Takeuchi N, Suzuki H and Yoshikawa N 2017 Measurement of low bit-error-rates of adiabatic quantum-flux-parametron logic using a superconductor voltage driver *Appl. Phys. Lett.* **110** 202601

[37] Yamanashi Y, Kainuma T, Yoshikawa N, Kataeva I, Akaike H, Fujimaki A, Tanaka M, Takagi N, Nagasawa S and Hidaka M 2010 100 GHz demonstrations based on the single-flux-quantum cell library for the 10 kA/cm$^2$ Nb multi-layer process *IEICE Trans. Electron.* **E93-C** 440–4